\documentclass[reprint,amsmath,amssymb,aps,twocolumn,superscriptaddress]{revtex4-1}

\usepackage{upgreek}
\usepackage{graphicx}
\usepackage{dcolumn}
\usepackage{bm}
\usepackage[caption=false]{subfig}
\usepackage{float}
\usepackage{placeins}
\usepackage{amsmath}
\usepackage{bm}
\usepackage{xcolor}
\usepackage{siunitx}
\usepackage{titlesec}
\usepackage{gensymb}
\usepackage[colorlinks=True,linkcolor=red,citecolor=blue,urlcolor=blue]{hyperref}
\renewcommand{\thesection}{\Roman{section}.} 
\newcommand {\bise}{Bi$_2$Se$_3$}

\titleformat{\section}
    {\normalfont\sc\filcenter}{\thesection}{1em}{}
\titlespacing*{\section}{0pt}{0.5\baselineskip}{\baselineskip}

\begin{document}

\preprint{APS/123-QED}

\title{Structural disorder-driven topological phase transition in noncentrosymmetric BiTeI}

\author
{Paul Corbae,$^{1,3}$ Frances Hellman$^{1,2,3}$ and Sin\'ead M. Griffin,$^{3,4}$\\
\normalsize{$^{1}$Department of Materials Science, University of California,}
\normalsize{Berkeley, California, 94720, USA}\\
\normalsize{$^{2}$Department of Physics, University of California,}
\normalsize{Berkeley, California, 94720, USA}\\
\normalsize{$^{3}$Materials Science Division, Lawrence Berkeley National Laboratory,}
\normalsize{Berkeley, California, 94720, USA}\\
\normalsize{$^{4}$
 Molecular Foundry, Lawrence Berkeley National Laboratory,}
\normalsize{Berkeley, California, 94720, USA}\\
\normalsize{$^\ast$To whom correspondence should be addressed; E-mail: pcorbae@berkeley.edu.}
}

\date{\today}

\begin{abstract}

We investigate the possibility of using structural disorder to induce a topological phase in a solid state system. Using first-principles calculations, we introduce structural disorder in the trivial insulator BiTeI and observe the emergence of a topological insulating phase. By modifying the bonding environments, the crystal-field splitting is enhanced and the spin-orbit interaction produces a band inversion in the bulk electronic structure. Analysis of the Wannier charge centers and the surface electronic structure reveals a strong topological insulator with Dirac surface states. Finally, we propose a prescription for inducing topological states from disorder in crystalline materials. Understanding how local environments produce topological phases is a key step for predicting disordered and amorphous topological materials. 

\end{abstract}

\maketitle

The discovery of nontrivial topological phases in materials systems has been at the forefront of condensed matter physics for both their fundamental importance and for their potential in applications ranging from low-power electronics to quantum computing. High-throughput searches and classification schemes for topological materials  exploit crystalline symmetries, and how these symmetries determine band connectivity \cite{Zhang:2019tp,Vergniory:2019ik,Tang:2019jx,Watanabe:2018bc,Bradlyn2017,Kruthoff_et_a:2017,Choudhary_et_al:2019}. Such symmetry indicators have enabled the prediction of thousands of crystalline topological materials. However, materials without intrinsic or long-range crystalline symmetries, such as amorphous and quasicrystalline materials \cite{Cain202015164,PhysRevLett.123.196401}, have no such classification scheme. Nonetheless, evidence for topological surface states has been experimentally and theoretically observed in an amorphous topologically insulating system \cite{corbae_et_al:2019,PhysRevLett.118.236402}. Recently, Marsal et al., \cite{adolfoquentin} exploited local environments to compute the topological phase diagram in a coordinated amorphous structure. 
These works find that the local interactions and connectivity determine the relevant features in the electronic structure to produce topological phases, and highlight the importance of local chemical environments in topological materials. Understanding how structural disorder contributes to and enhances topological phases will drive predictions of disordered and amorphous topological materials based on local structural properties.

Small gap semiconductors with a strong spin-orbit interaction (SOI) are ideal systems to probe how local environments affect the topology in the electronic structure. By tuning the chemical environment and interactions, the states near the Fermi level can produce a band inversion and topological phases \cite{Bahramy2012}. Systems that possess a strong SOI and lack a center of inversion are subject to Rashba spin-splitting. 
With no $I$-symmetry, the degeneracy between $\psi_{-k,\uparrow}$ and $\psi_{k,\uparrow}$ is lifted, the bands split at non-specific $k$-points and a band inversion can happen at points in the Brillouin zone (BZ) away from $\Gamma$. If the states near the Fermi level are of the same orbital character and also close in energy, they can couple together effectively through a Rashba Hamiltonian, further reducing the band gap at relevant points in the BZ \cite{PhysRevB.84.041202}. Therefore, a route to achieve topological states is to modify the chemical environment and crystal-field splitting (CFS) in strongly spin-orbit coupled materials without $I$-symmetry, such that the states near the Fermi level couple and produce a band inversion.

\begin{figure*}[!ht]
  \includegraphics[width=0.8\textwidth]{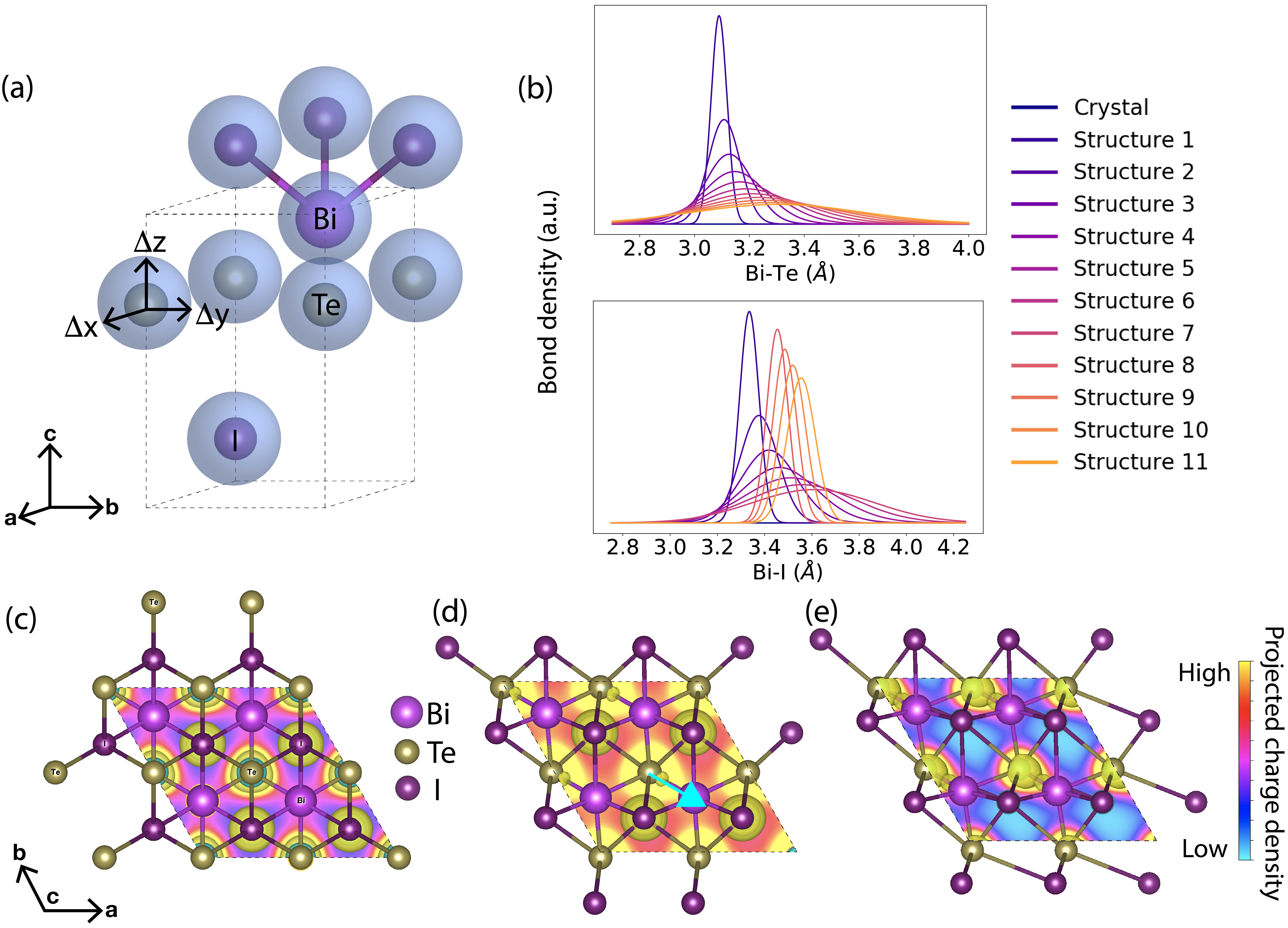}
  \caption{\textbf{Structural disorder induced charge redistribution}. (a) The BiTeI primitive unit cell. Blue spheres represent the allowed random displacements $\Delta x$, $\Delta y$, and $\Delta z$. (b) Normal distributions for the Bi-Te and Bi-I bond lengths. The Bi-Te bonds shift to a higher mean length and develop a larger $\sigma$. The Bi-I bonds initially move to a higher mean length with large $\sigma$ and then peak around \SI{3.50}{\AA} with low $\sigma$. (c-e) Partial charge density of the (001) plane for the bands near the Fermi level in structures with $d_{av}$=\SI{0.00}{\AA}, $d_{av}$=\SI{0.28}{\AA}, and $d_{av}$=\SI{0.56}{\AA}, respectively. As the structures become more disordered the charge density distorts into the $y$-direction (indicated by the blue arrow in (d)) and the charge moves to the Bi-Te bond. }
  \label{fig:structure}
\end{figure*}

 Several routes to inducing topological phase transitions (TPTs) from normal to topological insulator have been proposed in solid-state systems, including using pressure \cite{Mack9197} and strain \cite{PhysRevB.93.245303}. Topological Anderson insulators are an example where onsite disorder pushes a trivial insulator through a gapless state into a topologically nontrivial state \cite{PhysRevLett.103.196805}. However, no studies to date have looked at the effect of random structural disorder. 

In this work, we study how structural disorder affects the CFS and SOI in the small-gap semiconductor BiTeI and identify the important structural motifs that play a role in producing the nontrivial topology in the bandstructure. First, we present our scheme for structurally disordering BiTeI and the subsequent charge redistribution upon symmetry breaking. Using first-principles calculations we show that as bond lengths change, the CFS is enhanced leading to a reduction in the bandgap. We observe, with the addition of spin-orbit coupling (SOC), a TPT from trivial insulator to Weyl semimetal to topological insulator (TI) originating from a bulk band inversion. These topological phases emerge in a crystalline material with broken symmetries. This TPT is confirmed by studying the surface state spectrum and calculating topological invariants from Wannier charge centers. We observe a Dirac cone on the surface resulting from a strong $\mathbb{Z}_2$ index. This work provides an exciting pathway forward to understanding local environments in topological materials and their extension to amorphous systems, prompting a route for materials discovery \cite{corbae_et_al:2019}. 


BiTeI is a trigonal, noncentrosymmetric material adopting the $P3m1$ space group (156).  The primitive unit cell contains a single Bi, Te, and I atom configured in layers of triangular networks along the c-axis (Fig.~\ref{fig:structure}(a)), resulting in a $C_{3v}$ rotational symmetry about the $c$-axis. The Te-Bi-I network forms a trigonal prism surrounding the Bi atoms, which are separated by a van der Waals gap. In this undisordered structure the equilibrium Bi-Te bond length is \SI{3.07}{\AA} and the Bi-I bond length is \SI{3.30}{\AA}. In its crystalline form, it is a trivial insulator with a bandgap of 0.36 eV \cite{Lee_et_al:2011}. To understand how topology is influenced by the local chemical environment we generated disordered structures as follows: We pull a random number from a uniform distribution between \SI{-0.15}{\AA} and \SI{0.15}{\AA} and add it to each Wyckoff position in the unit cell for each direction. This generated a disordered structure with an average atomic displacement $d_{av}$=\SI{0.62}{\AA} per unit cell, similar in value to the change in interatomic spacing in amorphous \bise{} \cite{corbae_et_al:2019}. From this disordered structure, we created interpolated snapshots between the undisordered and fully disordered crystal with $d_{av}$=\SI{0.62}{\AA} to track the electronic and topological properties with increasing disorder. The final structure is in the $P1$ space group after the $C_{3v}$ and remaining symmetry elements are removed by the structural disorder. 

\begin{figure*}[!ht]
  \includegraphics[width=0.9\textwidth]{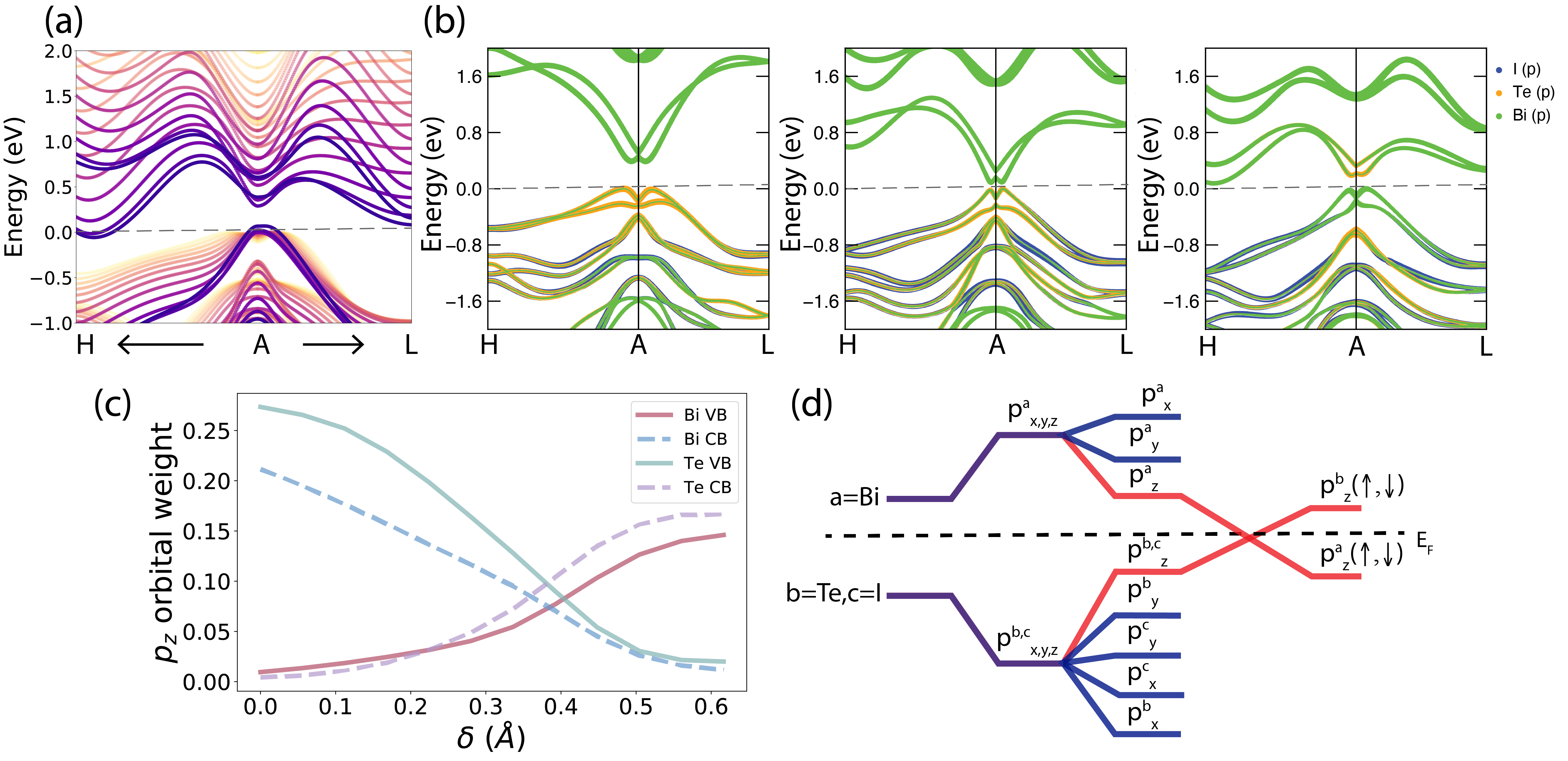}
  \caption{\textbf{Bulk electronic bandstructure}. (a) Bulk bandstructure in the $H-A-L$ direction without SOC. The structural disorder affects the CFS pushing the bands near the Fermi level closer and reducing the energy gap. Darker colors represnt more disordered structures. (b) The bandstructure with SOC for $d_{av}$=\SI{0.00}{\AA}, $d_{av}$=\SI{0.28}{\AA}, and $d_{av}$=\SI{0.45}{\AA}, respectively. There is a large Rashba splitting of the bands and a band inversion at the $A$ point due to the crystal field and SOI. The Bi weight in green moves frm the CB to VB and vice versa for Te in orange. (c) The Bi/Te $p_z$ orbital weight at $A$ as a function of average atomic displacement. With increasing displacement the Bi and Te weight of the VB (CB) switch around \SI{0.4}{\AA} implying a band inversion. (d) Energy level splitting diagram for disordered BiTeI after the TPT. The three splittings represent chemical bonding, the crystal field, and SOC. By breaking the $C_{3v}$ symmetry the $p_{x,y}$ orbitals are no longer degenerate.}
  \label{fig:bandstructure}
\end{figure*}

Fig.~\ref{fig:structure}(b) shows the normal distributions for Bi-Te and Bi-I bond lengths for our disordered structures. We see that with increasing disorder, the distribution of Bi-Te bond lengths develops a larger standard deviation while the mean bond length becomes larger. The smaller Bi-Te bonds shift to a value of \SI{2.9}{\AA} -- this Bi-Te bond plays an important role in the TPT as discussed later. Additionally, the distribution of Bi-I bond lengths spread out with large deviation and peak around \SI{3.5}{\AA} with lower deviation. 

The shortening of the Bi-Te bonds and the lengthening of the Bi-I bonds has important implications on the charge density. Fig.~\ref{fig:structure}(c) plots the charge density of the bands at the Fermi level near the $A$ point in the BZ involved in the TPT for the undisordered crystal. On the same plot, we also show the charge density for the bands projected onto the (001) plane which clearly shows the $C_{3v}$ symmetry and the charge sitting on the Te and I atoms. As we disorder the BiTeI structure, Fig.~\ref{fig:structure}(d-e), the charge density moves from being centered on the Te and I ions to the Bi-Te bond. This is important because, as we will describe later, the TPT band inversion occurs between the Bi and Te $p$ orbitals near the Fermi level. Furthermore, the charge density is redistributed in the $y$-direction due to the increased $p_y$ orbital presence at the Fermi level. This increase is attributed to the shortening of the Bi-Te bonds, which happens primarily in the $y$-direction. The $y$-direction is not specific, rather the important part is the shortening of the Bi-Te bond and the subsequent charge redistribution. In summary, in our disordered BiTeI structure, the Bi-Te bond gets shorter causing a charge redistribution along the Bi-Te bond, resulting in a broken 3-fold rotation symmetry and a new crystal-field environment for the states near the Fermi level. 

Next, we examine the influence of the structural changes on the electronic and topological properties of disordered BiTeI. Full DFT calculation details are given in the supplemental materials \cite{supp_bitei}. The calculated electronic structure is shown in Fig.~\ref{fig:bandstructure} with and without SOC included to disentangle  how crystal-field effects and SOC change with atomic displacement. Due to the lack of a center of inversion, we examine the states along the high-symmetry plane, $H-A-L$, to track topological band inversion. Fig.~\ref{fig:bandstructure}(a) shows the calculated electronic bandstructure for increasing values of $d_{av}$ without SOC. We find that for increasing $d_{av}$, the resulting changes in the CFS at $A$  reduces the band gap and pushes the bands at the Fermi level closer together. This large crystal-field enhancement is a result of the lifting of the degeneracy between the $p_x$ and $p_y$ orbitals upon displacement when the three-fold rotational symmetry is broken. This causes an increased $p_y$ orbital contribution near the Fermi level which is seen in the partial charge density. The orbital overlap produces large splitting pushing the states near the Fermi level closer together seen in Fig. \ref{fig:structure}(e) and Fig. \ref{fig:bandstructure}(a). 

\begin{figure}[!ht]
  \includegraphics[width=0.49\textwidth]{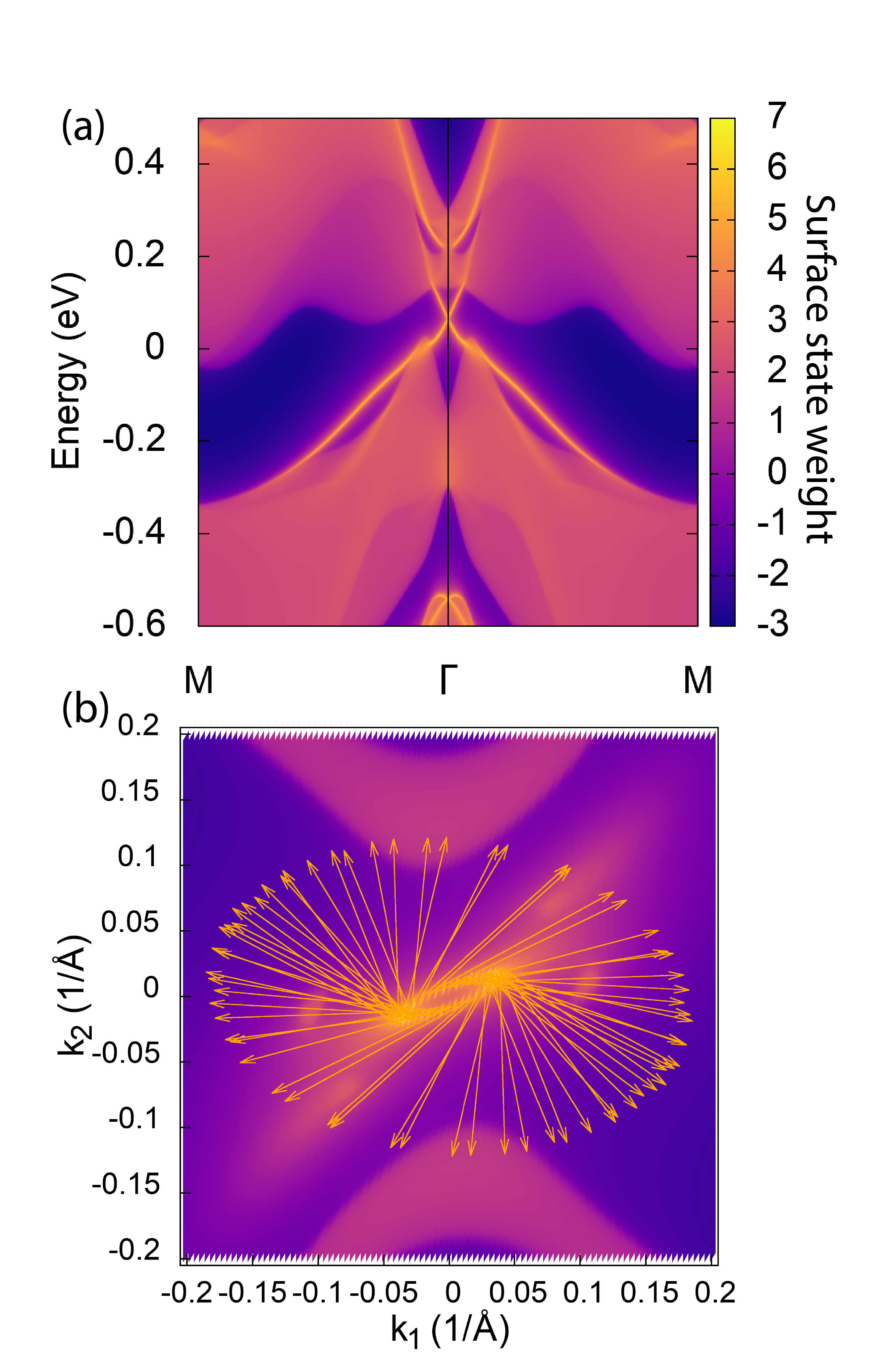}
  \caption{\textbf{Calculated surface state spectrum for disordered BiTeI}. (a) The momentum dependent local density of states shows a topological Dirac cone around the surface $\Gamma$ point. (b) The Fermi surface of the topological surface state. The Dirac cone is disordered due to the structural disorder. Arrows correspond to the spin texture of the Fermi surface. Brighter colors represent a higher local density of states.}
  \label{fig:topology}
\end{figure}

To understand the role of SOC on the electronic structure, Fig.~\ref{fig:bandstructure}(b) plots the bandstructure for $d_{av}=\SI{0.0}{\AA}$, $d_{av}=\SI{0.3}{\AA}$, and $d_{av}=\SI{0.5}{\AA}$, respectively, with SOC included. SOC breaks the spin degeneracy giving rise to splitting at positions in the Brillouin zone away from high-symmetry points and causes a large Rashba spin-splitting in the bulk bands near the Fermi level in all structures \cite{rashba}. Importantly, by incorporating SOC we observe a reduction of the bandgap, and, with increasing $d_{av}$, a band inversion occurs at the $A$ point. This band inversion produces the TPT in disordered BiTeI. The origin of the band inversion can be understood by considering the $p$-orbital projections of Bi, Te, and I onto the bandstructure. Initially the Te (orange) and I (blue) weight is concentrated in the valence band (VB) and the Bi weight (green) in the conduction band (CB). At the TPT the bands touch at the $A$ point, leading to a Weyl semimetal phase \cite{PhysRevB.90.155316}. After the transition the Bi weight is in the VB and the Te weight is in the CB, which is quantitatively shown in Fig.~\ref{fig:bandstructure}(c). The Bi and Te $p_z$ orbital weight, the wavefunction projected onto corresponding spherical harmonic, switch at $d_{av}\sim\SI{0.4}{\AA}$ with the Bi weight decreasing in the CB and the Te weight increasing (vice versa for the VB). Importantly, the average atomic displacement of \SI{0.4}{\AA} is observed in amorphous TI systems and so is a physically reasonable amount of disorder that could be induced in amorphous materials \cite{corbae_et_al:2019}. Additionally, the peak in the distribution of coordination numbers in the disordered structures moves from six to five exactly at the TPT, producing a new crystal field. The origin of the TPT can be traced back to the changing crystal field with disorder.

The bulk electronic structure results are summarized in Fig.~\ref{fig:bandstructure}(d). We find the states at the Fermi level, namely Bi-$p$ in the CB and Te/I-$p$ in the VB, dominate the TPT. As previously studies have shown, in crystalline BiTeI the in-plane  $C_{3v}$ rotational symmetry results in degenerate $p_{x,y}$ orbitals split in energy from the $p_z$ orbital due to the semi-ionic polar trigonal prismatic coordination of the Bi \cite{Ishizaka2011v2}. This results in Bi-$p_z$ and Te-$p_z$ states at the edge of the CB and VB respectively \cite{Zhang2009}, separated by a calculated band gap of \SI{0.38}{eV}. 
Introducing disorder causes the average bond lengths to change, notably shortening the Bi-Te bond length results in greater orbital overlap between the Bi-$p_z$ and Te-$p_z$ states, pushing the bands near the Fermi level closer together and reducing the bandgap. This disorder-induced reduction in bandgap is then sufficient, once SOC is included, to cause a band inversion of the two opposite polarity bands, resulting in a TPT to a topological insulating state. 

The bulk electronic band inversion at the $A$ point should manifest itself as a topological surface state  with a nontrivial topological invariant. Since the TPT is that of an ordinary insulator to a TI in a noncentrosymmetric crystal upon structural disorder, we calculate the $\mathbb{Z}_2$ topological invariant for six TR-invariant planes in the BZ \cite{PhysRevB.83.235401}. By tracking the evolution of the hybrid Wannier charge centers across these planes in the BZ we construct the topological index set \cite{supp_bitei}. After the TPT $\mathbb{Z}_2=1;\left(001\right)$ indicating the system is a strong TI. The strong topological index arising from the bulk electronic structure should manifest itself on the surface due to bulk-boundary correspondence. Fig.~\ref{fig:topology}(a) presents the surface state calculations performed on a slab of structurally disordered BiTeI after the TPT. The presence of a Dirac cone at the surface $\Gamma$ confirms the nontrivial bulk $\mathbb{Z}_2$ invariant, indicating with structural disorder we have produced a TI. The Dirac cone sits mid gap and passes through the top of the Rashba split VB as $\Gamma \rightarrow M$ before connecting to the VB at $M$. This unique surface spectrum allows for the interplay of bulk Rashba split bands with the topological surface state. In topological insulator, the surface spin-momentum locking with $H \propto \sigma \times \bm{k}$. The topological surface state's Fermi surface is shown in Fig.~\ref{fig:topology}(b). The annular structure is disordered in momentum space due to the structural disorder in real space. The spin texture is shown with arrows on the topological Fermi surface, confirming the spin-momentum locking. Increasing structural disorder produces a TPT leading to a strong TI with a spin-polarized Dirac cone.


The band inversion at the $A$ point of two Rashba split bands leads to the emergence of a Weyl semimetal topological phase when the bands touch \cite{PhysRevB.90.155316}. Calculating the Chern number for the $k_x=0$ plane gives $C=1$ meaning an intermediate Weyl phase in the structurally disordered BiTeI. Two Weyl nodes, which are sources/sinks of Berry curvature, are observed in the Berry curvature for the $k_x$ plane \cite{supp_bitei}. 

The local structure and effect of disorder play a key role in the TPT. We generated several sets of disordered interpolations via our random displacement process, different from the set presented in the main text. 
In the other sets of randomly disordered interpolations the bonds lengths develop different distributions leading to different local chemical environments and orbital overlaps. In one set of disordered interpolations the shortening of the Bi-Te bond causes a reduction in the band gap, but the crystal field splitting isn't enough to cause a band inversion, but larger amounts of disorder could potentially drive it past a TPT. The shortening of the Bi-Te bond and the corresponding charge redistribution was not present in another set of disordered interpolations, owing to the different local environments from nearest and next-nearest neighbor atoms seen in the radial distribution function \cite{supp_bitei}.
In fact, in these structures the CFS pushes the states near the Fermi level farther apart in energy resulting in a very large Rashba splitting in the bulk bands. In these non-topological structures $E_R = \SI{0.4}{eV}$ and the momentum offset $k_o=\SI{0.17}{\AA^{-1}}$ \cite{supp_bitei}, larger than any previously reported in literature \cite{Ishizaka2011}. These results demonstrate that structural disorder can generate both TPTs and colossal Rashba splitting in SOC materials making it a new tuning parameter for quantum properties in small gap semiconductors with large SOC.

Studying the influence of random structural disorder in SOC systems has implications for topological materials. From our results, we find that crystal-field engineering can be carried out with structural disorder as a new tuning parameter for topological phases. This leads us to a systematic prescription for inducing TPT in material as follows: (i) identify the orbital character of the states near the Fermi level, (ii) add random atomic disorder to break the degeneracy of the states and distort along the corresponding orbital direction to reduce the gap, (iii) close the gap and cause a band inversion. By incorporating tight-binding models and topological markers \cite{PhysRevResearch.2.013229,PhysRevLett.123.196401,adolfoquentin}, we can potentially discover many disordered and amorphous topological materials candidates. The topological phase in BiTeI could be realized in amorphous thin films grown via physical vapor deposition. Amorphous \bise{} showed an interatomic displacement of $\sim\SI{0.4}{\AA}$ when it was grown as a thin film \cite{corbae_et_al:2019}. Short range ordering coupled with the atomic displacement in disordered or amorphous BiTeI could perhaps show a topological phase in experiment.


In conclusion, we have found that by randomly introducing structural disorder a TPT from a normal insulator to topological insulator is achieved. Disordered BiTeI shows a bulk band inversion at the $A$ point in the BZ which manifests itself as a spin-polarized topological Dirac cone with a strong topological invariant. The physical mechanism for this is broken crystalline symmetries which produce a unique crystal-field environment that pushes the states near the Fermi level closer together. Our work is an effort to understand topological matter from a local bonding perspective and this study has implications for topological states that cannot be classified using crystalline symmetry indicators. Our results suggests a potential route to discover topological states in disordered and amorphous materials by identifying the local mechanisms (orbital inversions, etc.) which produce a TPT in the presence of disorder. Small-gap systems with strong SOC and well defined local environments, are exciting systems to study the interplay of structural disorder, symmetry breaking, and topology.

This work was primarily funded by the U.S. Department of Energy, Office of Science, Office of Basic Energy Sciences, Materials Sciences and Engineering Division under Contract No. DE-AC02-05-CH11231 (High-Coherence Multilayer Superconducting Structures for Large Scale Qubit Integration and Photonic Transduction program (QISLBNL). Work at the Molecular Foundry was supported by the Office of Science, Office of Basic Energy Sciences, of the U.S. Department of Energy under Contract No. DE-AC02-05CH11231. This research used resources of the National Energy Research Scientific Computing Center (NERSC), a U.S. Department of Energy Office of Science User Facility operated under Contract No. DE-AC02-05CH11231. This work was partially supported by the U.S. Department of Energy, Office of Science, Office of Basic Energy Sciences, Materials Sciences and Engineering Division under Contract No. DE-AC02-05- CH11231 within the Nonequilibrium Magnetic Materials Program (MSMAG). P. C. is supported by the National Science Foundation Graduate Research Fellowship under Grant No. 1752814.

\bibliography{bibbibbib.bib}

\begin{thebibliography}{31}%
\makeatletter
\providecommand \@ifxundefined [1]{%
 \@ifx{#1\undefined}
}%
\providecommand \@ifnum [1]{%
 \ifnum #1\expandafter \@firstoftwo
 \else \expandafter \@secondoftwo
 \fi
}%
\providecommand \@ifx [1]{%
 \ifx #1\expandafter \@firstoftwo
 \else \expandafter \@secondoftwo
 \fi
}%
\providecommand \natexlab [1]{#1}%
\providecommand \enquote  [1]{``#1''}%
\providecommand \bibnamefont  [1]{#1}%
\providecommand \bibfnamefont [1]{#1}%
\providecommand \citenamefont [1]{#1}%
\providecommand \href@noop [0]{\@secondoftwo}%
\providecommand \href [0]{\begingroup \@sanitize@url \@href}%
\providecommand \@href[1]{\@@startlink{#1}\@@href}%
\providecommand \@@href[1]{\endgroup#1\@@endlink}%
\providecommand \@sanitize@url [0]{\catcode `\\12\catcode `\$12\catcode
  `\&12\catcode `\#12\catcode `\^12\catcode `\_12\catcode `\%12\relax}%
\providecommand \@@startlink[1]{}%
\providecommand \@@endlink[0]{}%
\providecommand \url  [0]{\begingroup\@sanitize@url \@url }%
\providecommand \@url [1]{\endgroup\@href {#1}{\urlprefix }}%
\providecommand \urlprefix  [0]{URL }%
\providecommand \Eprint [0]{\href }%
\providecommand \doibase [0]{http://dx.doi.org/}%
\providecommand \selectlanguage [0]{\@gobble}%
\providecommand \bibinfo  [0]{\@secondoftwo}%
\providecommand \bibfield  [0]{\@secondoftwo}%
\providecommand \translation [1]{[#1]}%
\providecommand \BibitemOpen [0]{}%
\providecommand \bibitemStop [0]{}%
\providecommand \bibitemNoStop [0]{.\EOS\space}%
\providecommand \EOS [0]{\spacefactor3000\relax}%
\providecommand \BibitemShut  [1]{\csname bibitem#1\endcsname}%
\let\auto@bib@innerbib\@empty
\bibitem [{\citenamefont {Zhang}\ \emph {et~al.}(2019)\citenamefont {Zhang},
  \citenamefont {Jiang}, \citenamefont {Song}, \citenamefont {Huang},
  \citenamefont {He}, \citenamefont {Fang}, \citenamefont {Weng},\ and\
  \citenamefont {Fang}}]{Zhang:2019tp}%
  \BibitemOpen
  \bibfield  {author} {\bibinfo {author} {\bibfnamefont {T.}~\bibnamefont
  {Zhang}}, \bibinfo {author} {\bibfnamefont {Y.}~\bibnamefont {Jiang}},
  \bibinfo {author} {\bibfnamefont {Z.}~\bibnamefont {Song}}, \bibinfo {author}
  {\bibfnamefont {H.}~\bibnamefont {Huang}}, \bibinfo {author} {\bibfnamefont
  {Y.}~\bibnamefont {He}}, \bibinfo {author} {\bibfnamefont {Z.}~\bibnamefont
  {Fang}}, \bibinfo {author} {\bibfnamefont {H.}~\bibnamefont {Weng}}, \ and\
  \bibinfo {author} {\bibfnamefont {C.}~\bibnamefont {Fang}},\ }\href {\doibase
  10.1038/s41586-019-0944-6} {\bibfield  {journal} {\bibinfo  {journal}
  {Nature}\ }\textbf {\bibinfo {volume} {566}},\ \bibinfo {pages} {475}
  (\bibinfo {year} {2019})}\BibitemShut {NoStop}%
\bibitem [{\citenamefont {Vergniory}\ \emph {et~al.}(2019)\citenamefont
  {Vergniory}, \citenamefont {Elcoro}, \citenamefont {Felser}, \citenamefont
  {Regnault}, \citenamefont {Bernevig},\ and\ \citenamefont
  {Wang}}]{Vergniory:2019ik}%
  \BibitemOpen
  \bibfield  {author} {\bibinfo {author} {\bibfnamefont {M.~G.}\ \bibnamefont
  {Vergniory}}, \bibinfo {author} {\bibfnamefont {L.}~\bibnamefont {Elcoro}},
  \bibinfo {author} {\bibfnamefont {C.}~\bibnamefont {Felser}}, \bibinfo
  {author} {\bibfnamefont {N.}~\bibnamefont {Regnault}}, \bibinfo {author}
  {\bibfnamefont {B.~A.}\ \bibnamefont {Bernevig}}, \ and\ \bibinfo {author}
  {\bibfnamefont {Z.}~\bibnamefont {Wang}},\ }\href {\doibase
  10.1038/s41586-019-0954-4} {\bibfield  {journal} {\bibinfo  {journal} {Nature
  Publishing Group}\ }\textbf {\bibinfo {volume} {566}},\ \bibinfo {pages}
  {480} (\bibinfo {year} {2019})}\BibitemShut {NoStop}%
\bibitem [{\citenamefont {Tang}\ \emph {et~al.}(2019)\citenamefont {Tang},
  \citenamefont {Po}, \citenamefont {Vishwanath},\ and\ \citenamefont
  {Wan}}]{Tang:2019jx}%
  \BibitemOpen
  \bibfield  {author} {\bibinfo {author} {\bibfnamefont {F.}~\bibnamefont
  {Tang}}, \bibinfo {author} {\bibfnamefont {H.~C.}\ \bibnamefont {Po}},
  \bibinfo {author} {\bibfnamefont {A.}~\bibnamefont {Vishwanath}}, \ and\
  \bibinfo {author} {\bibfnamefont {X.}~\bibnamefont {Wan}},\ }\href {\doibase
  10.1038/s41586-019-0937-5} {\bibfield  {journal} {\bibinfo  {journal} {Nature
  Publishing Group}\ }\textbf {\bibinfo {volume} {566}},\ \bibinfo {pages}
  {486} (\bibinfo {year} {2019})}\BibitemShut {NoStop}%
\bibitem [{\citenamefont {Watanabe}\ \emph {et~al.}(2018)\citenamefont
  {Watanabe}, \citenamefont {Po},\ and\ \citenamefont
  {Vishwanath}}]{Watanabe:2018bc}%
  \BibitemOpen
  \bibfield  {author} {\bibinfo {author} {\bibfnamefont {H.}~\bibnamefont
  {Watanabe}}, \bibinfo {author} {\bibfnamefont {H.~C.}\ \bibnamefont {Po}}, \
  and\ \bibinfo {author} {\bibfnamefont {A.}~\bibnamefont {Vishwanath}},\
  }\href {\doibase 10.1126/sciadv.aat8685} {\bibfield  {journal} {\bibinfo
  {journal} {Science Advances}\ }\textbf {\bibinfo {volume} {4}},\ \bibinfo
  {pages} {eaat8685} (\bibinfo {year} {2018})}\BibitemShut {NoStop}%
\bibitem [{\citenamefont {{Bradlyn}}\ \emph {et~al.}(2017)\citenamefont
  {{Bradlyn}}, \citenamefont {{Elcoro}}, \citenamefont {{Cano}}, \citenamefont
  {{Vergniory}}, \citenamefont {{Wang}}, \citenamefont {{Felser}},
  \citenamefont {{Aroyo}},\ and\ \citenamefont {{Bernevig}}}]{Bradlyn2017}%
  \BibitemOpen
  \bibfield  {author} {\bibinfo {author} {\bibfnamefont {B.}~\bibnamefont
  {{Bradlyn}}}, \bibinfo {author} {\bibfnamefont {L.}~\bibnamefont {{Elcoro}}},
  \bibinfo {author} {\bibfnamefont {J.}~\bibnamefont {{Cano}}}, \bibinfo
  {author} {\bibfnamefont {M.~G.}\ \bibnamefont {{Vergniory}}}, \bibinfo
  {author} {\bibfnamefont {Z.}~\bibnamefont {{Wang}}}, \bibinfo {author}
  {\bibfnamefont {C.}~\bibnamefont {{Felser}}}, \bibinfo {author}
  {\bibfnamefont {M.~I.}\ \bibnamefont {{Aroyo}}}, \ and\ \bibinfo {author}
  {\bibfnamefont {B.~A.}\ \bibnamefont {{Bernevig}}},\ }\href {\doibase
  10.1038/nature23268} {\bibfield  {journal} {\bibinfo  {journal} {Nature}\
  }\textbf {\bibinfo {volume} {547}},\ \bibinfo {pages} {298} (\bibinfo {year}
  {2017})}\BibitemShut {NoStop}%
\bibitem [{\citenamefont {Kruthoff}\ \emph {et~al.}(2017)\citenamefont
  {Kruthoff}, \citenamefont {de~Boer}, \citenamefont {van Wezel}, \citenamefont
  {Kane},\ and\ \citenamefont {Slager}}]{Kruthoff_et_a:2017}%
  \BibitemOpen
  \bibfield  {author} {\bibinfo {author} {\bibfnamefont {J.}~\bibnamefont
  {Kruthoff}}, \bibinfo {author} {\bibfnamefont {J.}~\bibnamefont {de~Boer}},
  \bibinfo {author} {\bibfnamefont {J.}~\bibnamefont {van Wezel}}, \bibinfo
  {author} {\bibfnamefont {C.~L.}\ \bibnamefont {Kane}}, \ and\ \bibinfo
  {author} {\bibfnamefont {R.-J.}\ \bibnamefont {Slager}},\ }\href@noop {}
  {\bibfield  {journal} {\bibinfo  {journal} {Physical Review X}\ }\textbf
  {\bibinfo {volume} {7}},\ \bibinfo {pages} {041069} (\bibinfo {year}
  {2017})}\BibitemShut {NoStop}%
\bibitem [{\citenamefont {Choudhary}\ \emph {et~al.}(2019)\citenamefont
  {Choudhary}, \citenamefont {Garrity},\ and\ \citenamefont
  {Tavazza}}]{Choudhary_et_al:2019}%
  \BibitemOpen
  \bibfield  {author} {\bibinfo {author} {\bibfnamefont {K.}~\bibnamefont
  {Choudhary}}, \bibinfo {author} {\bibfnamefont {K.~F.}\ \bibnamefont
  {Garrity}}, \ and\ \bibinfo {author} {\bibfnamefont {F.}~\bibnamefont
  {Tavazza}},\ }\href@noop {} {\bibfield  {journal} {\bibinfo  {journal}
  {Scientific reports}\ }\textbf {\bibinfo {volume} {9}},\ \bibinfo {pages} {1}
  (\bibinfo {year} {2019})}\BibitemShut {NoStop}%
\bibitem [{\citenamefont {Cain}\ \emph {et~al.}(2020)\citenamefont {Cain},
  \citenamefont {Azizi}, \citenamefont {Conrad}, \citenamefont {Griffin},\ and\
  \citenamefont {Zettl}}]{Cain202015164}%
  \BibitemOpen
  \bibfield  {author} {\bibinfo {author} {\bibfnamefont {J.~D.}\ \bibnamefont
  {Cain}}, \bibinfo {author} {\bibfnamefont {A.}~\bibnamefont {Azizi}},
  \bibinfo {author} {\bibfnamefont {M.}~\bibnamefont {Conrad}}, \bibinfo
  {author} {\bibfnamefont {S.~M.}\ \bibnamefont {Griffin}}, \ and\ \bibinfo
  {author} {\bibfnamefont {A.}~\bibnamefont {Zettl}},\ }\href {\doibase
  10.1073/pnas.2015164117} {\bibfield  {journal} {\bibinfo  {journal}
  {Proceedings of the National Academy of Sciences}\ } (\bibinfo {year}
  {2020}),\ 10.1073/pnas.2015164117},\ \Eprint
  {http://arxiv.org/abs/https://www.pnas.org/content/early/2020/10/02/2015164117.full.pdf}
  {https://www.pnas.org/content/early/2020/10/02/2015164117.full.pdf}
  \BibitemShut {NoStop}%
\bibitem [{\citenamefont {Varjas}\ \emph {et~al.}(2019)\citenamefont {Varjas},
  \citenamefont {Lau}, \citenamefont {P\"oyh\"onen}, \citenamefont {Akhmerov},
  \citenamefont {Pikulin},\ and\ \citenamefont
  {Fulga}}]{PhysRevLett.123.196401}%
  \BibitemOpen
  \bibfield  {author} {\bibinfo {author} {\bibfnamefont {D.}~\bibnamefont
  {Varjas}}, \bibinfo {author} {\bibfnamefont {A.}~\bibnamefont {Lau}},
  \bibinfo {author} {\bibfnamefont {K.}~\bibnamefont {P\"oyh\"onen}}, \bibinfo
  {author} {\bibfnamefont {A.~R.}\ \bibnamefont {Akhmerov}}, \bibinfo {author}
  {\bibfnamefont {D.~I.}\ \bibnamefont {Pikulin}}, \ and\ \bibinfo {author}
  {\bibfnamefont {I.~C.}\ \bibnamefont {Fulga}},\ }\href {\doibase
  10.1103/PhysRevLett.123.196401} {\bibfield  {journal} {\bibinfo  {journal}
  {Phys. Rev. Lett.}\ }\textbf {\bibinfo {volume} {123}},\ \bibinfo {pages}
  {196401} (\bibinfo {year} {2019})}\BibitemShut {NoStop}%
\bibitem [{\citenamefont {Corbae}\ \emph {et~al.}(2019)\citenamefont {Corbae},
  \citenamefont {Ciocys},\ and\ \citenamefont {et~al.}}]{corbae_et_al:2019}%
  \BibitemOpen
  \bibfield  {author} {\bibinfo {author} {\bibfnamefont {P.}~\bibnamefont
  {Corbae}}, \bibinfo {author} {\bibfnamefont {S.}~\bibnamefont {Ciocys}}, \
  and\ \bibinfo {author} {\bibnamefont {et~al.}},\ }\href@noop {} {\bibfield
  {journal} {\bibinfo  {journal} {arXiv preprint arXiv:1910.13412}\ } (\bibinfo
  {year} {2019})}\BibitemShut {NoStop}%
\bibitem [{\citenamefont {Agarwala}\ and\ \citenamefont
  {Shenoy}(2017)}]{PhysRevLett.118.236402}%
  \BibitemOpen
  \bibfield  {author} {\bibinfo {author} {\bibfnamefont {A.}~\bibnamefont
  {Agarwala}}\ and\ \bibinfo {author} {\bibfnamefont {V.~B.}\ \bibnamefont
  {Shenoy}},\ }\href {\doibase 10.1103/PhysRevLett.118.236402} {\bibfield
  {journal} {\bibinfo  {journal} {Phys. Rev. Lett.}\ }\textbf {\bibinfo
  {volume} {118}},\ \bibinfo {pages} {236402} (\bibinfo {year}
  {2017})}\BibitemShut {NoStop}%
\bibitem [{\citenamefont {Marsal}\ \emph {et~al.}()\citenamefont {Marsal},
  \citenamefont {Varjas},\ and\ \citenamefont {Grushin}}]{adolfoquentin}%
  \BibitemOpen
  \bibfield  {author} {\bibinfo {author} {\bibfnamefont {Q.}~\bibnamefont
  {Marsal}}, \bibinfo {author} {\bibfnamefont {D.}~\bibnamefont {Varjas}}, \
  and\ \bibinfo {author} {\bibfnamefont {A.}~\bibnamefont {Grushin}},\ }\href
  {https://arxiv.org/pdf/2003.13701.pdf} {\bibinfo  {journal}
  {arXiv:2003.13701}\ }\BibitemShut {NoStop}%
\bibitem [{\citenamefont {Bahramy}\ \emph {et~al.}(2012)\citenamefont
  {Bahramy}, \citenamefont {Yang}, \citenamefont {Arita},\ and\ \citenamefont
  {Nagaosa}}]{Bahramy2012}%
  \BibitemOpen
\bibfield  {journal} {  }\bibfield  {author} {\bibinfo {author} {\bibfnamefont
  {M.~S.}\ \bibnamefont {Bahramy}}, \bibinfo {author} {\bibfnamefont {B.-J.}\
  \bibnamefont {Yang}}, \bibinfo {author} {\bibfnamefont {R.}~\bibnamefont
  {Arita}}, \ and\ \bibinfo {author} {\bibfnamefont {N.}~\bibnamefont
  {Nagaosa}},\ }\href {\doibase 10.1038/ncomms1679} {\bibfield  {journal}
  {\bibinfo  {journal} {Nature Communications}\ }\textbf {\bibinfo {volume}
  {3}},\ \bibinfo {pages} {679} (\bibinfo {year} {2012})}\BibitemShut {NoStop}%
\bibitem [{\citenamefont {Bahramy}\ \emph {et~al.}(2011)\citenamefont
  {Bahramy}, \citenamefont {Arita},\ and\ \citenamefont
  {Nagaosa}}]{PhysRevB.84.041202}%
  \BibitemOpen
  \bibfield  {author} {\bibinfo {author} {\bibfnamefont {M.~S.}\ \bibnamefont
  {Bahramy}}, \bibinfo {author} {\bibfnamefont {R.}~\bibnamefont {Arita}}, \
  and\ \bibinfo {author} {\bibfnamefont {N.}~\bibnamefont {Nagaosa}},\ }\href
  {\doibase 10.1103/PhysRevB.84.041202} {\bibfield  {journal} {\bibinfo
  {journal} {Phys. Rev. B}\ }\textbf {\bibinfo {volume} {84}},\ \bibinfo
  {pages} {041202(R)} (\bibinfo {year} {2011})}\BibitemShut {NoStop}%
\bibitem [{\citenamefont {Mack}\ \emph {et~al.}(2019)\citenamefont {Mack},
  \citenamefont {Griffin},\ and\ \citenamefont {Neaton}}]{Mack9197}%
  \BibitemOpen
  \bibfield  {author} {\bibinfo {author} {\bibfnamefont {S.~A.}\ \bibnamefont
  {Mack}}, \bibinfo {author} {\bibfnamefont {S.~M.}\ \bibnamefont {Griffin}}, \
  and\ \bibinfo {author} {\bibfnamefont {J.~B.}\ \bibnamefont {Neaton}},\
  }\href {\doibase 10.1073/pnas.1821533116} {\bibfield  {journal} {\bibinfo
  {journal} {Proceedings of the National Academy of Sciences}\ }\textbf
  {\bibinfo {volume} {116}},\ \bibinfo {pages} {9197} (\bibinfo {year}
  {2019})},\ \Eprint
  {http://arxiv.org/abs/https://www.pnas.org/content/116/19/9197.full.pdf}
  {https://www.pnas.org/content/116/19/9197.full.pdf} \BibitemShut {NoStop}%
\bibitem [{\citenamefont {Zhang}\ \emph {et~al.}(2016)\citenamefont {Zhang},
  \citenamefont {Xie}, \citenamefont {Cai}, \citenamefont {Zhang},
  \citenamefont {Ma}, \citenamefont {Chen}, \citenamefont {Zhu}, \citenamefont
  {Hu},\ and\ \citenamefont {Zeng}}]{PhysRevB.93.245303}%
  \BibitemOpen
  \bibfield  {author} {\bibinfo {author} {\bibfnamefont {S.}~\bibnamefont
  {Zhang}}, \bibinfo {author} {\bibfnamefont {M.}~\bibnamefont {Xie}}, \bibinfo
  {author} {\bibfnamefont {B.}~\bibnamefont {Cai}}, \bibinfo {author}
  {\bibfnamefont {H.}~\bibnamefont {Zhang}}, \bibinfo {author} {\bibfnamefont
  {Y.}~\bibnamefont {Ma}}, \bibinfo {author} {\bibfnamefont {Z.}~\bibnamefont
  {Chen}}, \bibinfo {author} {\bibfnamefont {Z.}~\bibnamefont {Zhu}}, \bibinfo
  {author} {\bibfnamefont {Z.}~\bibnamefont {Hu}}, \ and\ \bibinfo {author}
  {\bibfnamefont {H.}~\bibnamefont {Zeng}},\ }\href {\doibase
  10.1103/PhysRevB.93.245303} {\bibfield  {journal} {\bibinfo  {journal} {Phys.
  Rev. B}\ }\textbf {\bibinfo {volume} {93}},\ \bibinfo {pages} {245303}
  (\bibinfo {year} {2016})}\BibitemShut {NoStop}%
\bibitem [{\citenamefont {Groth}\ \emph {et~al.}(2009)\citenamefont {Groth},
  \citenamefont {Wimmer}, \citenamefont {Akhmerov}, \citenamefont
  {Tworzyd\l{}o},\ and\ \citenamefont {Beenakker}}]{PhysRevLett.103.196805}%
  \BibitemOpen
  \bibfield  {author} {\bibinfo {author} {\bibfnamefont {C.~W.}\ \bibnamefont
  {Groth}}, \bibinfo {author} {\bibfnamefont {M.}~\bibnamefont {Wimmer}},
  \bibinfo {author} {\bibfnamefont {A.~R.}\ \bibnamefont {Akhmerov}}, \bibinfo
  {author} {\bibfnamefont {J.}~\bibnamefont {Tworzyd\l{}o}}, \ and\ \bibinfo
  {author} {\bibfnamefont {C.~W.~J.}\ \bibnamefont {Beenakker}},\ }\href
  {\doibase 10.1103/PhysRevLett.103.196805} {\bibfield  {journal} {\bibinfo
  {journal} {Phys. Rev. Lett.}\ }\textbf {\bibinfo {volume} {103}},\ \bibinfo
  {pages} {196805} (\bibinfo {year} {2009})}\BibitemShut {NoStop}%
\bibitem [{\citenamefont {Lee}\ \emph {et~al.}(2011)\citenamefont {Lee},
  \citenamefont {Schober}, \citenamefont {Bahramy}, \citenamefont {Murakawa},
  \citenamefont {Onose}, \citenamefont {Arita}, \citenamefont {Nagaosa},\ and\
  \citenamefont {Tokura}}]{Lee_et_al:2011}%
  \BibitemOpen
  \bibfield  {author} {\bibinfo {author} {\bibfnamefont {J.~S.}\ \bibnamefont
  {Lee}}, \bibinfo {author} {\bibfnamefont {G.~A.~H.}\ \bibnamefont {Schober}},
  \bibinfo {author} {\bibfnamefont {M.~S.}\ \bibnamefont {Bahramy}}, \bibinfo
  {author} {\bibfnamefont {H.}~\bibnamefont {Murakawa}}, \bibinfo {author}
  {\bibfnamefont {Y.}~\bibnamefont {Onose}}, \bibinfo {author} {\bibfnamefont
  {R.}~\bibnamefont {Arita}}, \bibinfo {author} {\bibfnamefont
  {N.}~\bibnamefont {Nagaosa}}, \ and\ \bibinfo {author} {\bibfnamefont
  {Y.}~\bibnamefont {Tokura}},\ }\href {\doibase
  10.1103/PhysRevLett.107.117401} {\bibfield  {journal} {\bibinfo  {journal}
  {Phys. Rev. Lett.}\ }\textbf {\bibinfo {volume} {107}},\ \bibinfo {pages}
  {117401} (\bibinfo {year} {2011})}\BibitemShut {NoStop}%
\bibitem [{sup(2020)}]{supp_bitei}%
  \BibitemOpen
  \href@noop {} {\bibfield  {journal} {\bibinfo  {journal} {See Supplemental
  Material for calculation details, discussion of the structure, WCC evolution,
  and Rashba/Weyl effects.}\ } (\bibinfo {year} {2020})}\BibitemShut {NoStop}%
\bibitem [{\citenamefont {Rashba}(1960)}]{rashba}%
  \BibitemOpen
  \bibfield  {author} {\bibinfo {author} {\bibfnamefont {E.~I.}\ \bibnamefont
  {Rashba}},\ }\href@noop {} {\bibfield  {journal} {\bibinfo  {journal} {Sov.
  Phys. Solid State}\ }\textbf {\bibinfo {volume} {2}},\ \bibinfo {pages}
  {1109–1122} (\bibinfo {year} {1960})}\BibitemShut {NoStop}%
\bibitem [{\citenamefont {Liu}\ and\ \citenamefont
  {Vanderbilt}(2014)}]{PhysRevB.90.155316}%
  \BibitemOpen
  \bibfield  {author} {\bibinfo {author} {\bibfnamefont {J.}~\bibnamefont
  {Liu}}\ and\ \bibinfo {author} {\bibfnamefont {D.}~\bibnamefont
  {Vanderbilt}},\ }\href {\doibase 10.1103/PhysRevB.90.155316} {\bibfield
  {journal} {\bibinfo  {journal} {Phys. Rev. B}\ }\textbf {\bibinfo {volume}
  {90}},\ \bibinfo {pages} {155316} (\bibinfo {year} {2014})}\BibitemShut
  {NoStop}%
\bibitem [{\citenamefont {Ishizaka}\ and\ \citenamefont {et.
  al}(2011)}]{Ishizaka2011v2}%
  \BibitemOpen
  \bibfield  {author} {\bibinfo {author} {\bibfnamefont {K.}~\bibnamefont
  {Ishizaka}}\ and\ \bibinfo {author} {\bibnamefont {et. al}},\ }\href
  {\doibase 10.1038/nmat3051} {\bibfield  {journal} {\bibinfo  {journal}
  {Nature Materials}\ }\textbf {\bibinfo {volume} {10}},\ \bibinfo {pages}
  {521} (\bibinfo {year} {2011})}\BibitemShut {NoStop}%
\bibitem [{\citenamefont {Zhang}\ \emph {et~al.}(2009)\citenamefont {Zhang},
  \citenamefont {Liu}, \citenamefont {Qi}, \citenamefont {Dai}, \citenamefont
  {Fang},\ and\ \citenamefont {Zhang}}]{Zhang2009}%
  \BibitemOpen
  \bibfield  {author} {\bibinfo {author} {\bibfnamefont {H.}~\bibnamefont
  {Zhang}}, \bibinfo {author} {\bibfnamefont {C.-X.}\ \bibnamefont {Liu}},
  \bibinfo {author} {\bibfnamefont {X.-L.}\ \bibnamefont {Qi}}, \bibinfo
  {author} {\bibfnamefont {X.}~\bibnamefont {Dai}}, \bibinfo {author}
  {\bibfnamefont {Z.}~\bibnamefont {Fang}}, \ and\ \bibinfo {author}
  {\bibfnamefont {S.-C.}\ \bibnamefont {Zhang}},\ }\href {\doibase
  10.1038/nphys1270} {\bibfield  {journal} {\bibinfo  {journal} {Nature
  Physics}\ }\textbf {\bibinfo {volume} {5}},\ \bibinfo {pages} {438} (\bibinfo
  {year} {2009})}\BibitemShut {NoStop}%
\bibitem [{\citenamefont {Soluyanov}\ and\ \citenamefont
  {Vanderbilt}(2011)}]{PhysRevB.83.235401}%
  \BibitemOpen
  \bibfield  {author} {\bibinfo {author} {\bibfnamefont {A.~A.}\ \bibnamefont
  {Soluyanov}}\ and\ \bibinfo {author} {\bibfnamefont {D.}~\bibnamefont
  {Vanderbilt}},\ }\href {\doibase 10.1103/PhysRevB.83.235401} {\bibfield
  {journal} {\bibinfo  {journal} {Phys. Rev. B}\ }\textbf {\bibinfo {volume}
  {83}},\ \bibinfo {pages} {235401} (\bibinfo {year} {2011})}\BibitemShut
  {NoStop}%
\bibitem [{\citenamefont {Ishizaka}\ \emph {et~al.}(2011)\citenamefont
  {Ishizaka}, \citenamefont {Bahramy}, \citenamefont {Murakawa}, \citenamefont
  {Sakano}, \citenamefont {Shimojima}, \citenamefont {Sonobe}, \citenamefont
  {Koizumi}, \citenamefont {Shin}, \citenamefont {Miyahara}, \citenamefont
  {Kimura}, \citenamefont {Miyamoto}, \citenamefont {Okuda}, \citenamefont
  {Namatame}, \citenamefont {Taniguchi}, \citenamefont {Arita}, \citenamefont
  {Nagaosa}, \citenamefont {Kobayashi}, \citenamefont {Murakami}, \citenamefont
  {Kumai}, \citenamefont {Kaneko}, \citenamefont {Onose},\ and\ \citenamefont
  {Tokura}}]{Ishizaka2011}%
  \BibitemOpen
  \bibfield  {author} {\bibinfo {author} {\bibfnamefont {K.}~\bibnamefont
  {Ishizaka}}, \bibinfo {author} {\bibfnamefont {M.~S.}\ \bibnamefont
  {Bahramy}}, \bibinfo {author} {\bibfnamefont {H.}~\bibnamefont {Murakawa}},
  \bibinfo {author} {\bibfnamefont {M.}~\bibnamefont {Sakano}}, \bibinfo
  {author} {\bibfnamefont {T.}~\bibnamefont {Shimojima}}, \bibinfo {author}
  {\bibfnamefont {T.}~\bibnamefont {Sonobe}}, \bibinfo {author} {\bibfnamefont
  {K.}~\bibnamefont {Koizumi}}, \bibinfo {author} {\bibfnamefont
  {S.}~\bibnamefont {Shin}}, \bibinfo {author} {\bibfnamefont {H.}~\bibnamefont
  {Miyahara}}, \bibinfo {author} {\bibfnamefont {A.}~\bibnamefont {Kimura}},
  \bibinfo {author} {\bibfnamefont {K.}~\bibnamefont {Miyamoto}}, \bibinfo
  {author} {\bibfnamefont {T.}~\bibnamefont {Okuda}}, \bibinfo {author}
  {\bibfnamefont {H.}~\bibnamefont {Namatame}}, \bibinfo {author}
  {\bibfnamefont {M.}~\bibnamefont {Taniguchi}}, \bibinfo {author}
  {\bibfnamefont {R.}~\bibnamefont {Arita}}, \bibinfo {author} {\bibfnamefont
  {N.}~\bibnamefont {Nagaosa}}, \bibinfo {author} {\bibfnamefont
  {K.}~\bibnamefont {Kobayashi}}, \bibinfo {author} {\bibfnamefont
  {Y.}~\bibnamefont {Murakami}}, \bibinfo {author} {\bibfnamefont
  {R.}~\bibnamefont {Kumai}}, \bibinfo {author} {\bibfnamefont
  {Y.}~\bibnamefont {Kaneko}}, \bibinfo {author} {\bibfnamefont
  {Y.}~\bibnamefont {Onose}}, \ and\ \bibinfo {author} {\bibfnamefont
  {Y.}~\bibnamefont {Tokura}},\ }\href {\doibase 10.1038/nmat3051} {\bibfield
  {journal} {\bibinfo  {journal} {Nature Materials}\ }\textbf {\bibinfo
  {volume} {10}},\ \bibinfo {pages} {521} (\bibinfo {year} {2011})}\BibitemShut
  {NoStop}%
\bibitem [{\citenamefont {Varjas}\ \emph {et~al.}(2020)\citenamefont {Varjas},
  \citenamefont {Fruchart}, \citenamefont {Akhmerov},\ and\ \citenamefont
  {Perez-Piskunow}}]{PhysRevResearch.2.013229}%
  \BibitemOpen
  \bibfield  {author} {\bibinfo {author} {\bibfnamefont {D.}~\bibnamefont
  {Varjas}}, \bibinfo {author} {\bibfnamefont {M.}~\bibnamefont {Fruchart}},
  \bibinfo {author} {\bibfnamefont {A.~R.}\ \bibnamefont {Akhmerov}}, \ and\
  \bibinfo {author} {\bibfnamefont {P.~M.}\ \bibnamefont {Perez-Piskunow}},\
  }\href {\doibase 10.1103/PhysRevResearch.2.013229} {\bibfield  {journal}
  {\bibinfo  {journal} {Phys. Rev. Research}\ }\textbf {\bibinfo {volume}
  {2}},\ \bibinfo {pages} {013229} (\bibinfo {year} {2020})}\BibitemShut
  {NoStop}%
\bibitem [{\citenamefont {Kresse}\ and\ \citenamefont
  {Furthm\"uller}(1996)}]{VASP1}%
  \BibitemOpen
  \bibfield  {author} {\bibinfo {author} {\bibfnamefont {G.}~\bibnamefont
  {Kresse}}\ and\ \bibinfo {author} {\bibfnamefont {J.}~\bibnamefont
  {Furthm\"uller}},\ }\href {\doibase 10.1103/PhysRevB.54.11169} {\bibfield
  {journal} {\bibinfo  {journal} {Phys. Rev. B}\ }\textbf {\bibinfo {volume}
  {54}},\ \bibinfo {pages} {11169} (\bibinfo {year} {1996})}\BibitemShut
  {NoStop}%
\bibitem [{\citenamefont {Kresse}\ and\ \citenamefont {Hafner}(1993)}]{VASP2}%
  \BibitemOpen
  \bibfield  {author} {\bibinfo {author} {\bibfnamefont {G.}~\bibnamefont
  {Kresse}}\ and\ \bibinfo {author} {\bibfnamefont {J.}~\bibnamefont
  {Hafner}},\ }\href@noop {} {\bibfield  {journal} {\bibinfo  {journal} {Phys.
  Rev. B}\ }\textbf {\bibinfo {volume} {48}},\ \bibinfo {pages} {13115}
  (\bibinfo {year} {1993})}\BibitemShut {NoStop}%
\bibitem [{\citenamefont {Perdew}\ \emph {et~al.}(1996)\citenamefont {Perdew},
  \citenamefont {Burke},\ and\ \citenamefont {Ernzerhof}}]{PBE1}%
  \BibitemOpen
  \bibfield  {author} {\bibinfo {author} {\bibfnamefont {J.~P.}\ \bibnamefont
  {Perdew}}, \bibinfo {author} {\bibfnamefont {K.}~\bibnamefont {Burke}}, \
  and\ \bibinfo {author} {\bibfnamefont {M.}~\bibnamefont {Ernzerhof}},\ }\href
  {\doibase 10.1103/PhysRevLett.77.3865} {\bibfield  {journal} {\bibinfo
  {journal} {Phys. Rev. Lett.}\ }\textbf {\bibinfo {volume} {77}},\ \bibinfo
  {pages} {3865} (\bibinfo {year} {1996})}\BibitemShut {NoStop}%
\bibitem [{\citenamefont {Mostofi}\ \emph {et~al.}(2014)\citenamefont
  {Mostofi}, \citenamefont {Yates}, \citenamefont {Pizzi}, \citenamefont {Lee},
  \citenamefont {Souza}, \citenamefont {Vanderbilt},\ and\ \citenamefont
  {Marzari}}]{MOSTOFI20142309}%
  \BibitemOpen
  \bibfield  {author} {\bibinfo {author} {\bibfnamefont {A.~A.}\ \bibnamefont
  {Mostofi}}, \bibinfo {author} {\bibfnamefont {J.~R.}\ \bibnamefont {Yates}},
  \bibinfo {author} {\bibfnamefont {G.}~\bibnamefont {Pizzi}}, \bibinfo
  {author} {\bibfnamefont {Y.-S.}\ \bibnamefont {Lee}}, \bibinfo {author}
  {\bibfnamefont {I.}~\bibnamefont {Souza}}, \bibinfo {author} {\bibfnamefont
  {D.}~\bibnamefont {Vanderbilt}}, \ and\ \bibinfo {author} {\bibfnamefont
  {N.}~\bibnamefont {Marzari}},\ }\href {\doibase
  https://doi.org/10.1016/j.cpc.2014.05.003} {\bibfield  {journal} {\bibinfo
  {journal} {Computer Physics Communications}\ }\textbf {\bibinfo {volume}
  {185}},\ \bibinfo {pages} {2309 } (\bibinfo {year} {2014})}\BibitemShut
  {NoStop}%
\bibitem [{\citenamefont {Wu}\ \emph {et~al.}(2018)\citenamefont {Wu},
  \citenamefont {Zhang}, \citenamefont {Song}, \citenamefont {Troyer},\ and\
  \citenamefont {Soluyanov}}]{WU2017}%
  \BibitemOpen
  \bibfield  {author} {\bibinfo {author} {\bibfnamefont {Q.}~\bibnamefont
  {Wu}}, \bibinfo {author} {\bibfnamefont {S.}~\bibnamefont {Zhang}}, \bibinfo
  {author} {\bibfnamefont {H.-F.}\ \bibnamefont {Song}}, \bibinfo {author}
  {\bibfnamefont {M.}~\bibnamefont {Troyer}}, \ and\ \bibinfo {author}
  {\bibfnamefont {A.~A.}\ \bibnamefont {Soluyanov}},\ }\href {\doibase
  https://doi.org/10.1016/j.cpc.2017.09.033} {\bibfield  {journal} {\bibinfo
  {journal} {Computer Physics Communications}\ }\textbf {\bibinfo {volume}
  {224}},\ \bibinfo {pages} {405 } (\bibinfo {year} {2018})}\BibitemShut
  {NoStop}%
\end{thebibliography}%

\renewcommand{\thefigure}{S\arabic{figure}}

\section*{Supplementary materials}

\subsection*{First principles calculations}

Our electronic structure calculations were performed using Density Functional Theory (DFT) with the projector augmented wave (PAW) formalism in the Vienna ab initio Simulation Package (VASP)\cite{VASP1, VASP2}. The exchange-correlation potentials were treated in the framework of generalized gradient approximation (GGA) of Perdew-Burke-Ernzerbof (PBE)\cite{PBE1}. Bi (6s, 6p), Te (5s,5p), and I (5s, 5p) electrons were treated as valence, and their wavefunctions expanded in plane waves to an energy cutoff of 600 eV.  A k-point grid of 6x6x4 with Gamma sampling was used. Spin-orbit coupling was added self-consistently.

\subsection*{Topological invariants and surface states}

Based on our first-principles results, we construct a Wannier function based tight-binding model using the WANNIER90 code \cite{MOSTOFI20142309}. We include the Bi, Te, and I $p$ orbitals in the model. We then calculated the topological invariants for different structures using the WANNIERTOOLS package \cite{WU2017}.

\section{Supplementary Text}

\subsection*{Structure}

To track the topological properties of BiTeI, 11 disordered structures were produced by randomly displacing the atoms from their equilibrium positions. 
The reason for the band inversion and topological phase transition is the shortening of the Bi-Te bond and subsequent charge redistribution to this bond due to the new crystal-field environment. 
The Bi and Te atoms get closer together in the $x,y$ directions shortening the bond. This is the reason for a larger $p_{x,y}$ orbital contribution at the Fermi level seen in Fig. S1.

\subsection*{Topological invariants}

The WCC for the $k_{3}=0.5$ plane is shown in Fig. S2, there is clearly a single crossing of the line cut with a WCC giving a $\mathbb{Z}_2$ index of 1. In conjunction with all the other planes, this shows the system is in a topologically insulating phase. 

\subsection*{Non-topological BiTeI}

Different realizations of structural disorder in BiTeI lead to non-topological systems. The radial distribution function is shown in Fig. S3. for different realizations with similar average displacement. The topological structure has peaks in the RDF at $\sim2.5\AA$ and $\sim2.8\AA$. The different local environment leads to a different modifications of crystal-field environment that push the bands near the Fermi level farther apart. This is shown in Fig. S4, the energy gap at the $A$ point in the BZ is larger than it is in the undistorted crystal. This different crystal field results from random atomic displacements that do not produce the relevant orbital overlaps required for topology (such as the $p_z$ orbital overlap in systems studied in the main text). This can be seen in the Bi, Te, and I orbital projections, these states reside in the same bands as the non-topological crystal. Although the system doesn't undergo a topological phase transformation it develops an even greater Rashba splitting seen in Fig. S4. The conduction band is split about the $A$ point, and the momentum offset is much greater than that of the pristine crystal. 

\subsection*{Intermediate Weyl Semimetal phase}

These topological phases have been discussed before as an intermediate phases between a trivial $\mathbb{Z}_2$ insulator and a nontrivial $\mathbb{Z}_2$ insulator which lacks a center of inversion \cite{PhysRevB.90.155316}. 
As BiTeI becomes more structurally disordered, the crystal-field splitting increases and the bands approach a band inversion. When the valence and conduction bands touch at the $A$ point in the BZ, a pair of Weyl nodes should emerge. Since pairs Weyl nodes are sources and sinks of Berry curvature, we should observe this phase by looking at the real part of $\Omega$. Fig. S5 shows the real part of $\Omega_x$ ($x$-component of the Berry curvature). There are two nodes in the BZ, each with opposite Berry curvature from the other (red vs. blue). This, in addition to the non-zero Chern number, signifies the presence of a Weyl phase as BiTeI trasnitions from a trivial insulator to topological insulator.

\section{Supplementary Figures}

\begin{figure*}
  \includegraphics[width=1\textwidth]{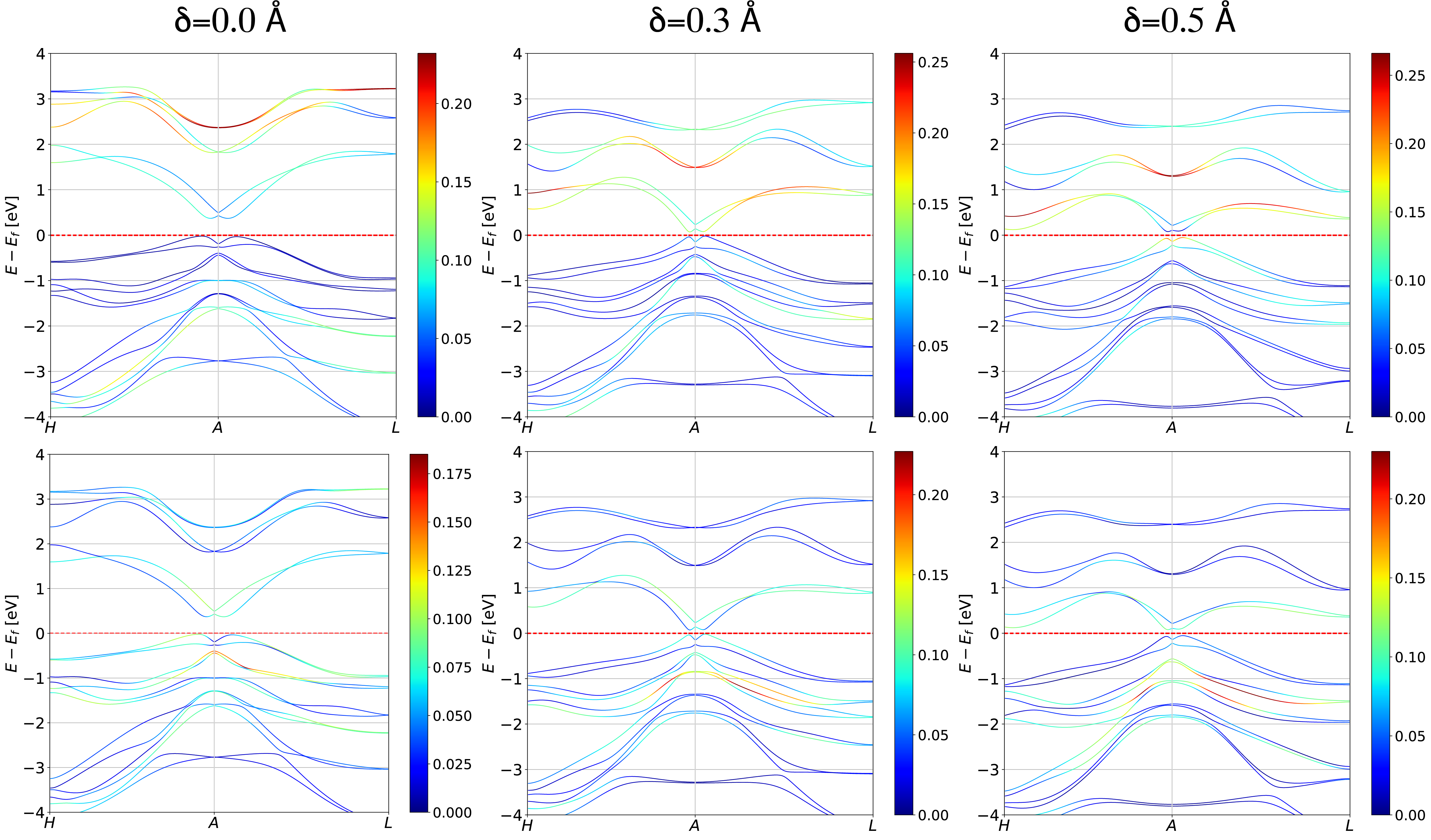}
  \caption{$p_y$ orbital projected bandstructure for three distortion magnitudes. The Fermi level is at 0 eV in each plot and marked with a line. Top row: Bi $p_y$ orbital. Bottom row: Te $p_y$ orbital. Color bar indicates the $p_y$ orbital weight, red specifies larger orbital weight and blue specifies low orbital weight.}
\end{figure*}

\begin{figure*}
    \includegraphics[width=1\textwidth]{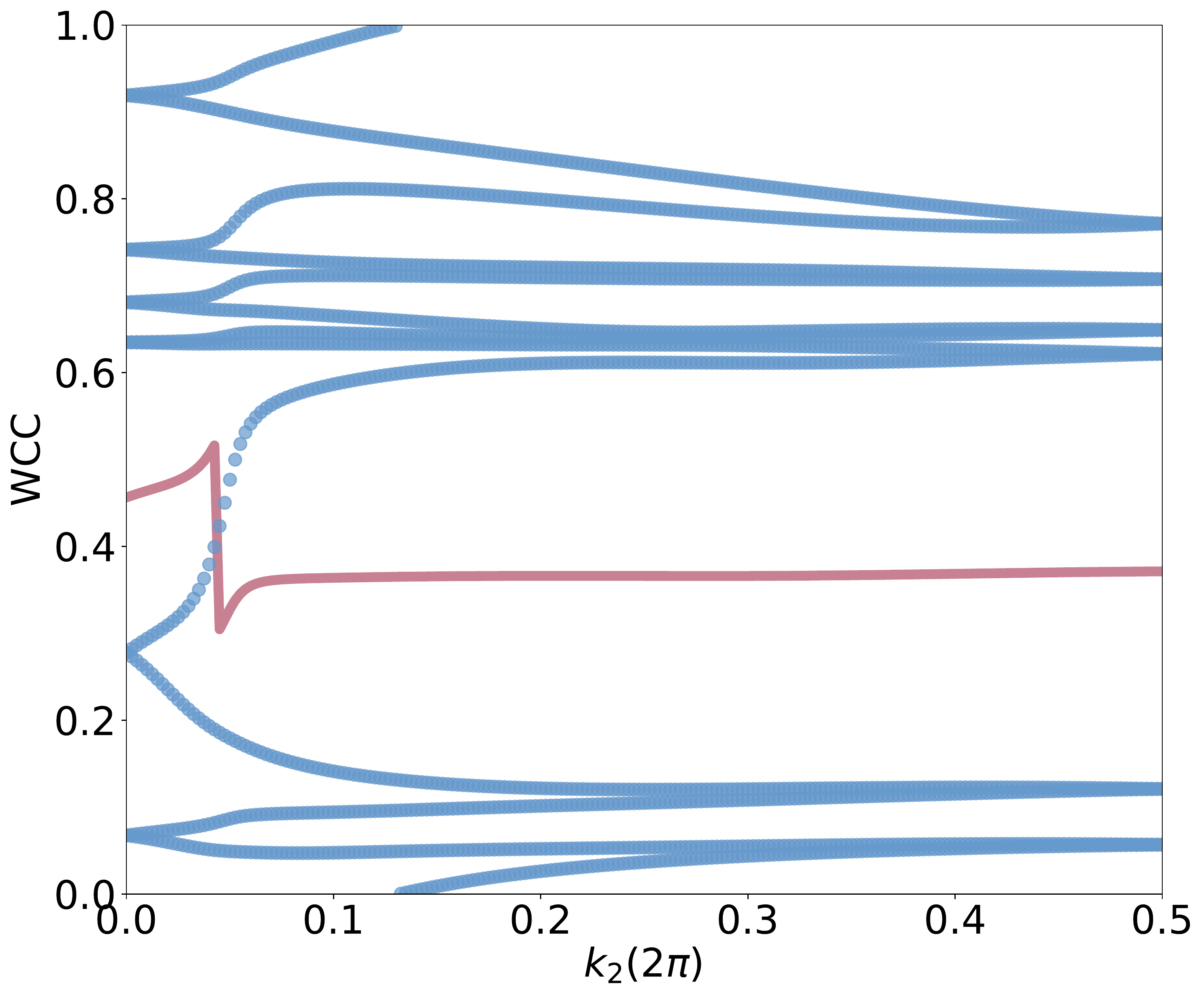}
    \caption{Hybrid WCC evolution for the $k_{3}=0.5$ plane. The red line is the largest gap between WCC’s and the blue lines are the WCC's. There is a single crosses indicating this plane has $\mathbb{Z}_2=1$ and the system is topologically insulating.}
\end{figure*}

\begin{figure*}
    \includegraphics[width=1\textwidth]{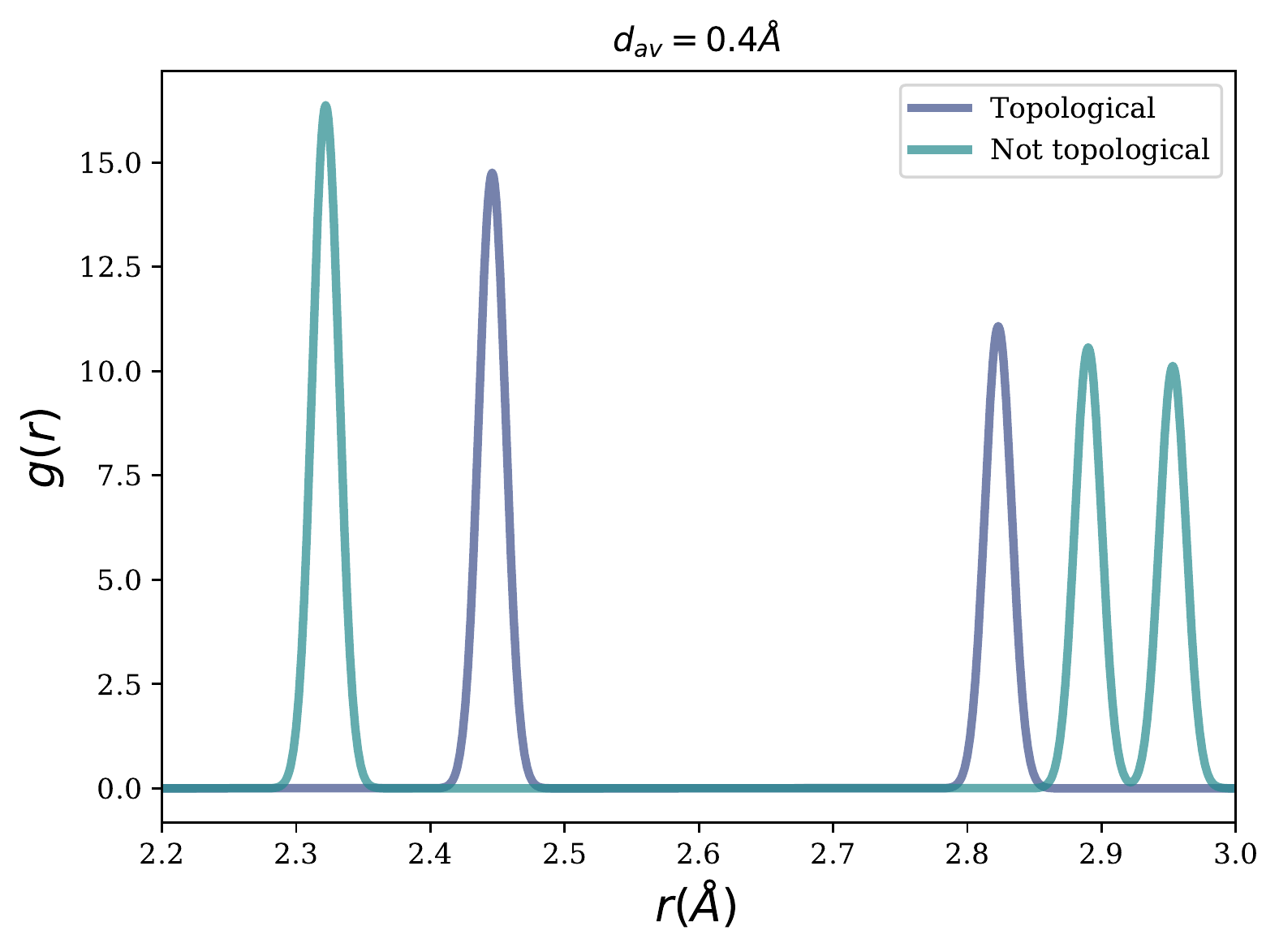}
        \caption{Computed RDF for different realizations of structural disorder in BiTeI. 
        The structure labeled not topoligcal has $d_{av}=\SI{0.4}{\AA}$ but does not possess the required Bi-Te bond shortening and charge redistribution for a topological phase transition.}
\end{figure*}

\begin{figure*}
  \includegraphics[width=1\textwidth]{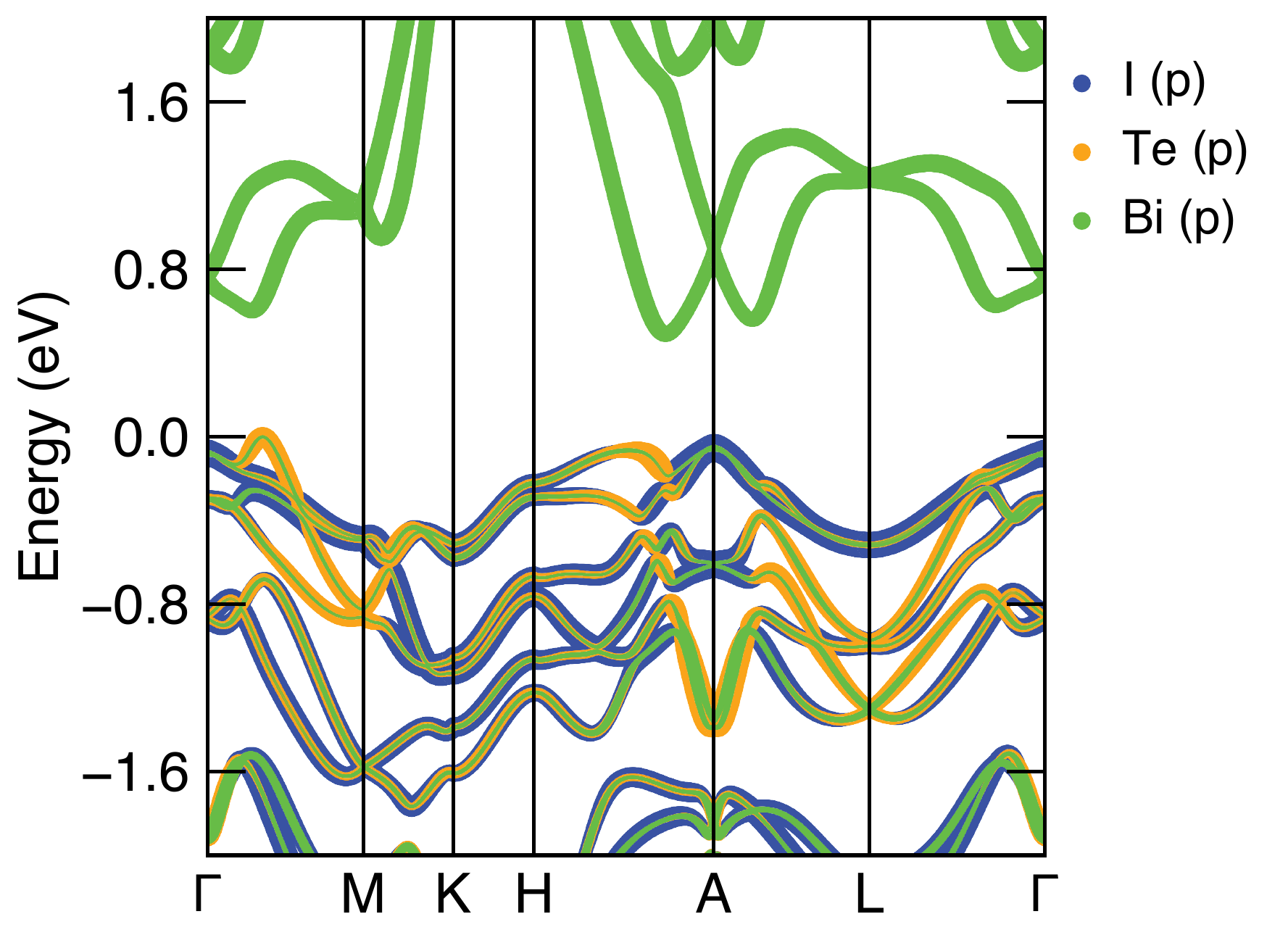}
  \caption{Electronic bandstructure for non-topological BiTeI with orbital projections with the Fermi level set to 0 eV. As the structural disorder becomes larger the Rashba splitting is greatly increased. The conduction band splitting at the $A$ point has $E_R = \SI{0.4}{eV}$ and the momentum offset $k_o=\SI{0.17}{\AA^{-1}}$.}
\end{figure*}

\begin{figure*}
  \includegraphics[width=1\textwidth]{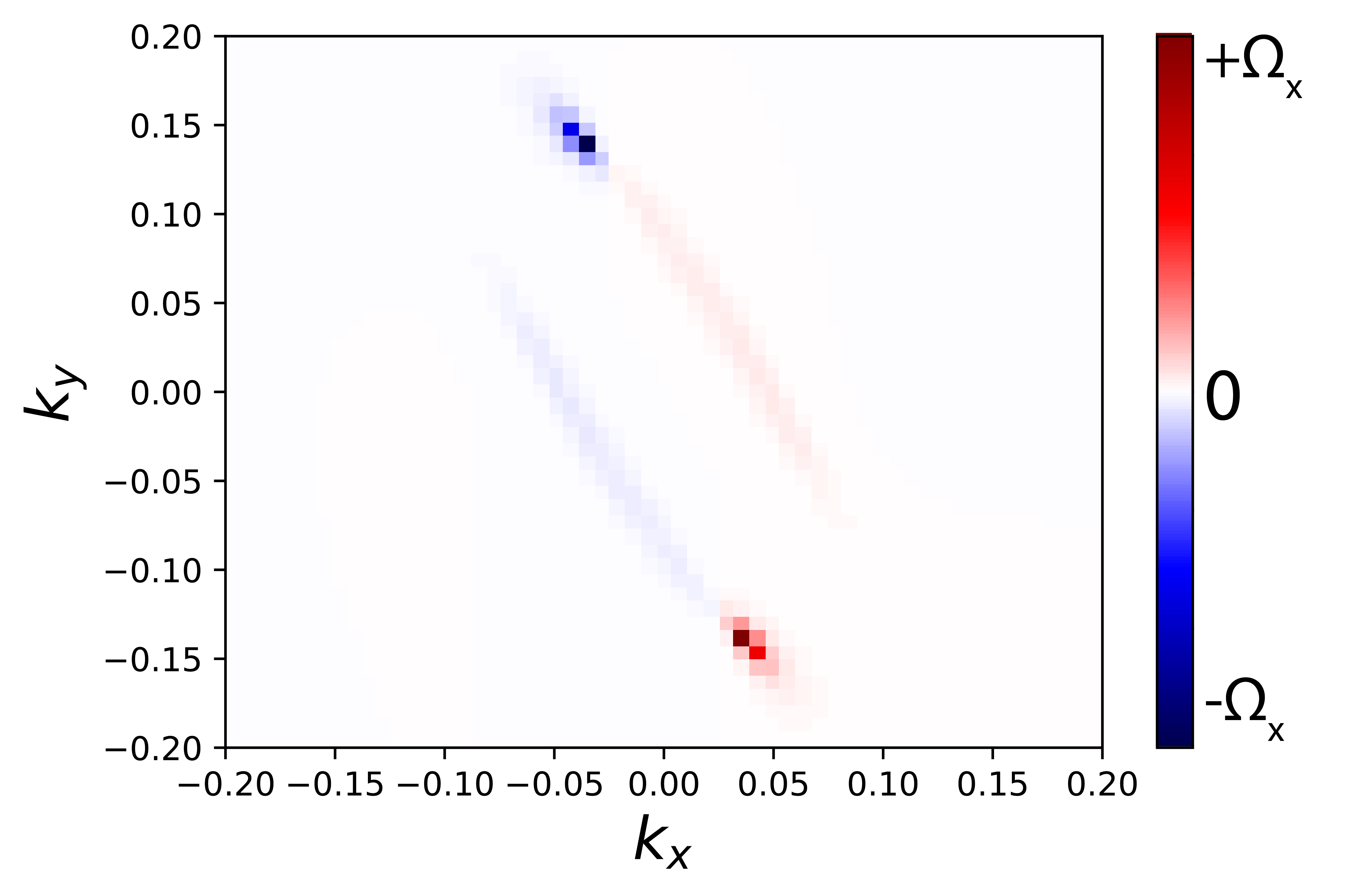}
  \caption{Real $\Omega_x$ at the topological phase transition. Two nodes with opposite Berry curvature are observed in the BZ at the Fermi level. Red and blue denote opposite Berry Curvature.}
\end{figure*}

\end{document}